\documentclass[journal]{IEEEtran}  %

\usepackage{multirow}     %
\usepackage{booktabs}     %
\usepackage{amsmath,amssymb,amsfonts}
\usepackage{rotating}
\usepackage{graphicx}
\usepackage{algorithm}
\usepackage{algorithmic}
\usepackage{adjustbox}
\usepackage[graphicx]{realboxes}

\usepackage{cite}
\usepackage{float}
\usepackage{textcomp}
\usepackage{xcolor}
\begin{document}

\bstctlcite{IEEEexample:BSTcontrol}
\title{Computing-In-Memory Aware Model Adaption For Edge Devices}

\author{\IEEEauthorblockN{Ming-Han Lin, and Tian-Sheuan Chang, \textit{Senior Member, IEEE}}
\thanks{This work was supported by the National Science and Technology Council, Taiwan, under Grant 111-2622-8-A49-018-SB, 110-2221-E-A49-148-MY3, 113-2221-E-A49-078-MY3, and 113-2640-E-A49-005.. The authors are affiliated with the Institute of Electronics, National Yang Ming Chiao Tung University, Taiwan. (e-mail: hanklin09@gmail.com, tschang@nycu.edu.tw). cited as: M.-H. Lin and T. S. Chang, "Computing-in-memory aware model adaption for edge devices", to be published in IEEE Transactions on Circuits and Systems for Artificial Intelligence, 2026. }%
\thanks{Manuscript received XXXX XX, 2025; revised XXXX XX, XXXX.}
}
\maketitle

\begin{abstract}%

Computing-in-Memory (CIM) macros have gained popularity for deep learning acceleration due to their highly parallel computation and low power consumption. However, limited macro size and ADC precision introduce throughput and accuracy bottlenecks. This paper proposes a two-stage CIM-aware model adaptation process. The first stage compresses the model and reallocates resources based on layer importance and macro size constraints, reducing model weight loading latency while improving resource utilization and maintaining accuracy. The second stage performs quantization-aware training, incorporating partial sum quantization and ADC precision to mitigate quantization errors in inference.
The proposed approach enhances CIM array utilization to 90\%, enables concurrent activation of up to 256 word lines, and achieves up to 93\% compression, all while preserving accuracy comparable to previous methods.

~
\end{abstract}

\section{Introduction}
\label{sec:introduction}

The proliferation of complex deep learning models has spurred the development of specialized hardware accelerators for edge devices, where power and latency are critical constraints. Computing-in-Memory (CIM) has emerged as a highly promising architecture, offering massive parallelism and reduced data movement by performing computations directly within the memory array. However, the practical deployment of CIM is hindered by two fundamental and interconnected challenges rooted in its physical limitations.

First, \textit{Hardware Mapping and Throughput Bottlenecks} arise from the constrained physical size of CIM macros. Modern deep neural networks are often too large to be stored entirely on-chip, necessitating that model weights be repeatedly loaded from off-chip memory. This frequent reloading incurs significant latency and energy overhead, negating many of CIM's intrinsic benefits.

Second, \textit{Computational Fidelity and Accuracy Degradation} are direct consequences of the precision-limited analog-to-digital converters (ADCs) inherent to CIM design. When convolutions are segmented due to hardware size limits, multiple analog partial sums are generated. Each of these sums must be quantized by the ADC, causing quantization errors to accumulate and severely degrade model accuracy. A common workaround is to severely restrict the number of concurrently activated wordlines to match ADC precision (e.g., activating only 16 wordlines for a 4-bit ADC). However, this drastically underutilizes the available parallelism of the CIM array and throttles performance.

To overcome these obstacles, researchers have proposed various model adaptation strategies. One line of work focuses on \textit{CIM-aware model compression and architecture search}. For instance, E-UPQ \cite{chang2023upq} enhances model sparsity through pruning and mixed-precision quantization but suffers from low macro utilization. XPert \cite{moitra2023xpert} co-searches for the neural architecture and peripheral circuits, but its rigid optimization constraints can limit flexibility. Similarly, CIMNet \cite{chen2024cimnet} uses a device-aware accuracy predictor for neural architecture search but overlooks the significant performance penalty caused by weight reloading.

Another line of work targets \textit{mitigating ADC quantization effects}. These methods aim to increase the effective number of bits (ENOB) by mapping the multiply-accumulate (MAC) distribution to the ADC's input range. Approaches include optimizing quantization ranges based on MAC statistics \cite{9626555}, using input-conditioned subrange reduction techniques \cite{10411993}, or learning analog scaling factors \cite{bai2023partial, kim2022extreme}. While effective, these methods often do not account for the large number of partial sums generated when many wordlines are activated in parallel, or are designed for smaller CIM macros \cite{bai2023partial}.

The existing literature reveals a critical gap: a holistic approach that simultaneously optimizes the model architecture for dense mapping onto the CIM array while also making the model inherently robust to the partial sum quantization errors that arise from maximizing parallelism. To bridge this gap, this paper proposes a tailored model adaptation method that adjusts the model architecture and recalibrates weights to mitigate quantization errors. Our approach reallocates limited resources, such as bitlines per convolutional layer, to enhance efficiency while maintaining or improving accuracy. We implement a two-stage quantization-aware training process that quantizes both weights and partial sums, simulating CIM behavior and reducing the impact of quantization on model accuracy.

The rest of the paper is organized as follows: Section II details the proposed methods, Section III presents the experimental results, and Section IV concludes the paper.

\section{Proposed CIM-Aware Model Adaption}
\label{session:Proposed Methods}

\subsection{The Target Multibit CIM Architecture}

\begin{figure}[h]
\centering
\includegraphics[width=0.75\linewidth]{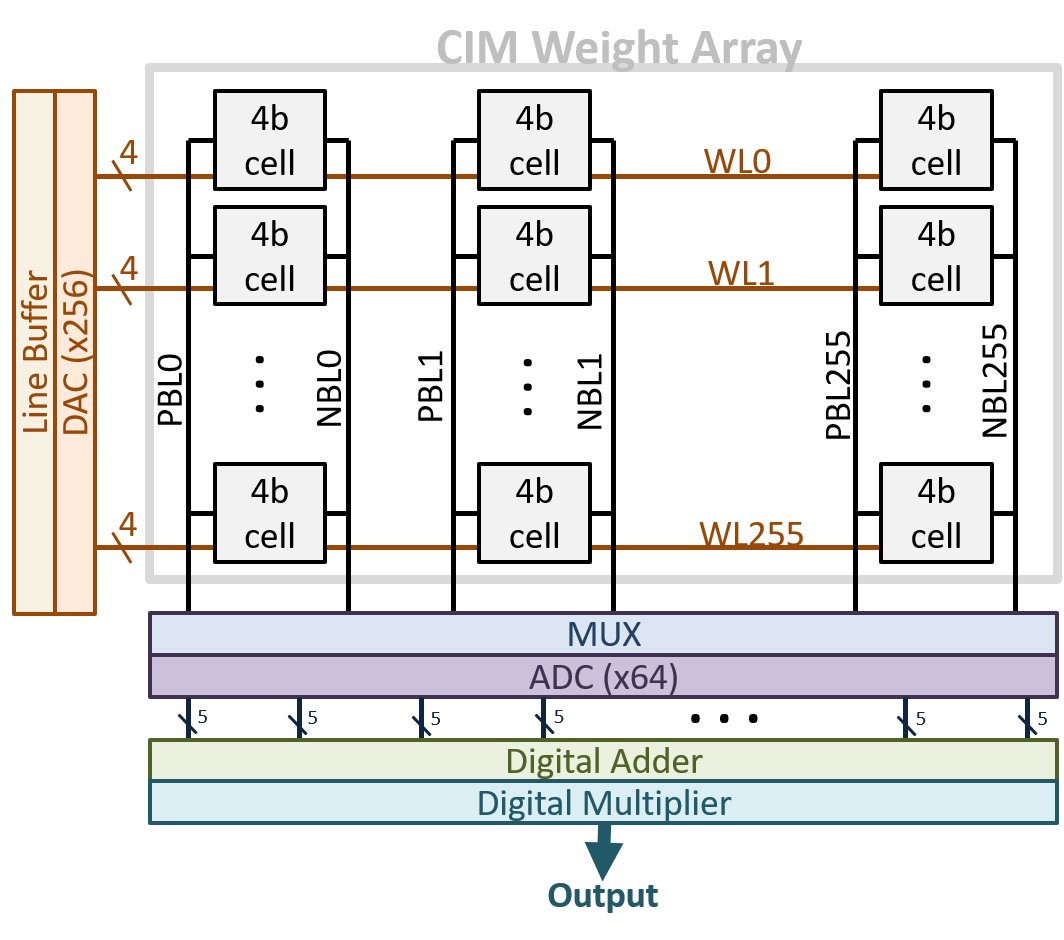}
\caption{4-bit CIM macro architecture}
\label{CIM_architecture}
\end{figure}

Fig.~\ref{CIM_architecture} illustrates the configuration of the CIM macro used in this paper. The workflow involves the following steps: a line buffer transfers 4-bit input data to a Digital-to-Analog Converter (DAC), converting it into an analog signal that enters the CIM weight array's wordlines. Each weight cell multiplies the input data, and the products are accumulated in each bitline. A multiplexer selects the processed signals, which are then converted into 5-bit digital partial sums by an ADC.

In terms of precision, each weight cell uses 4 bits, with parallel inputs converted to voltage by the DAC. The ADC then transforms the analog signal into a 5-bit digital format. This system requires only one ADC conversion for the multiply-accumulate operation, reducing the number of conversions by a factor of 16 compared to a bit-by-bit method, which helps minimize quantization errors, especially in the most significant bits (MSB).

The CIM array consists of 256 wordlines and 256 bitlines, along with 64 ADCs. Each weight cell stores 4 bits of data. The bitlines include positive (PBL) and negative (NBL) lines. The multiplexer selects different bitlines, and the ADCs operate in rotation to convert the analog signals into digital sums.

\begin{figure}[H]
\centering
\includegraphics[width=0.75\linewidth]{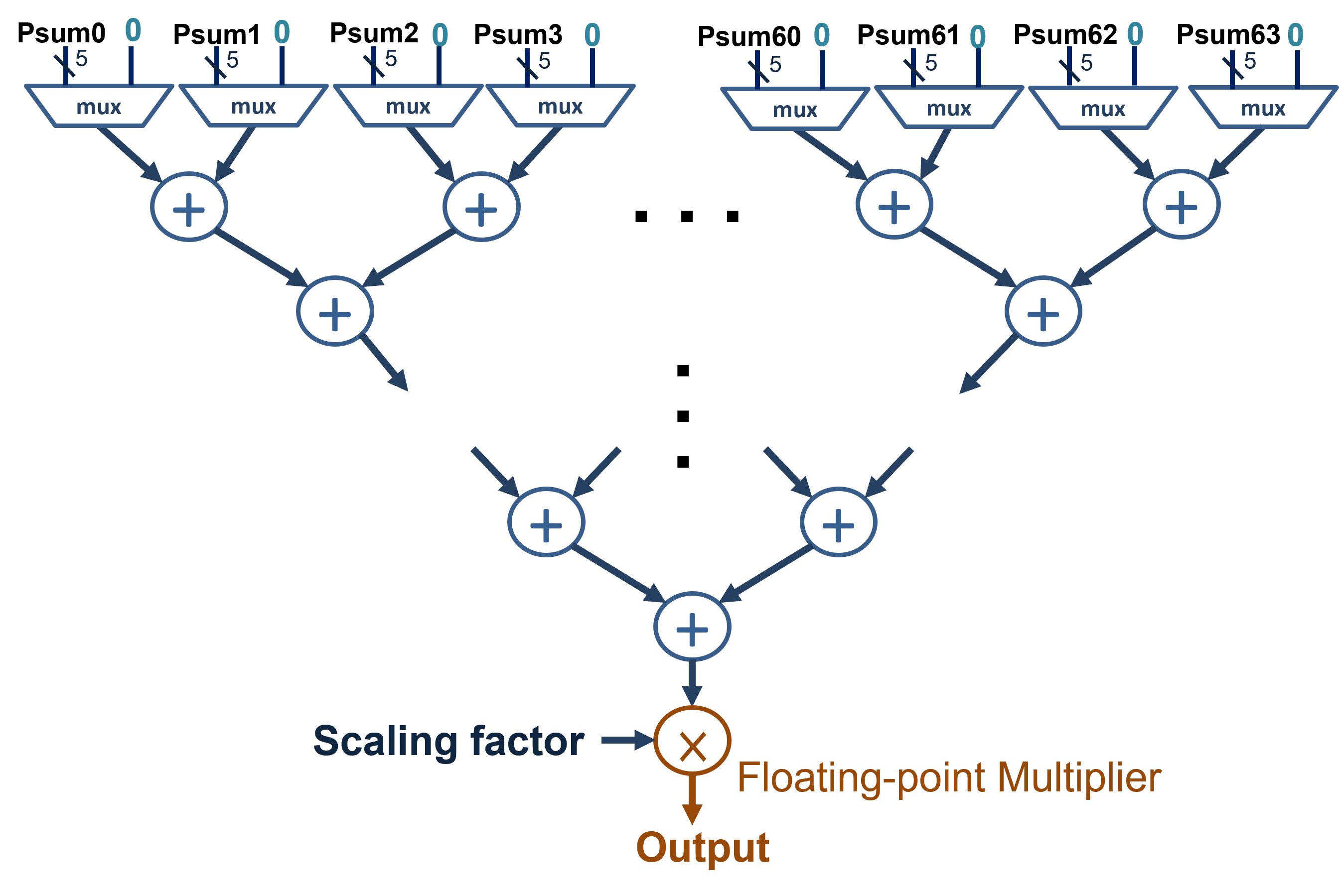}
\caption{The digital circuits that assist our CIM macro}
\label{digital_circuit}
\end{figure}

In Fig.~\ref{digital_circuit}, 64 5-bit partial sums are accumulated using an adder tree and then multiplied by a scaling factor. Since the 64 ADCs are not used simultaneously, a multiplexer at each ADC output selects the appropriate ADC for accumulation. The final scaling factor combines both the weight scaling factor and the ADC step size, addressing the need to reverse the effects of scaling. This is necessary because the weights, initially in decimal form, are quantized into 4-bit integers, and the partial sums from the ADC also undergo scaling during conversion.

\begin{figure}[H]
\centering
\includegraphics[width=1.0\linewidth]{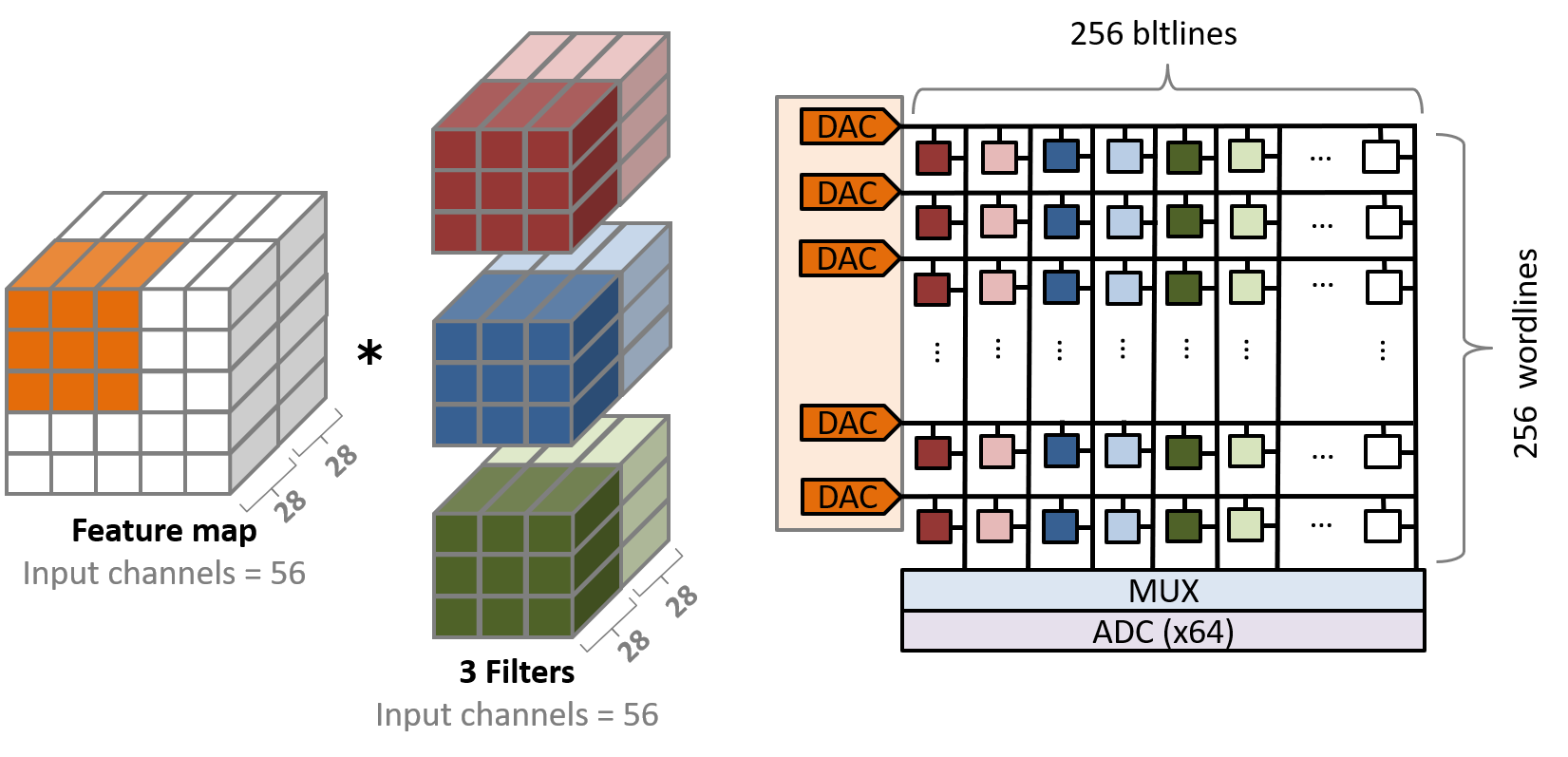}
\caption{Mapping convolution weights into a CIM macro}
\label{weight_mapping}
\end{figure}

Fig.~\ref{weight_mapping} illustrates the weight mapping for convolution. Due to the limited number of wordlines in the memory array, the multiply-accumulate operation cannot be completed in a single pass. Instead, the convolution kernel is divided into multiple parts based on the number of wordlines, processed in batches, and accumulated for the final result. For instance, with 256 wordlines and a 3x3 filter size, one bitline can handle up to 28 input channels, necessitating that any excess data be placed in the next bitline.

In the example, three filters are split into two parts, indicated by different colors, and stored in separate bitlines. The DAC inputs to the CIM macro include the orange section of the feature map, representing the first half of the input channels, which perform dot products with the corresponding darker sections of the filters. Consequently, only outputs from three bitlines are valid at this stage, while the remaining data will be processed subsequently.

\subsection{Overall Two-Stage Model Adaption Flow for CIM}

\begin{figure}[H]
\centering
\includegraphics[width=1\linewidth]{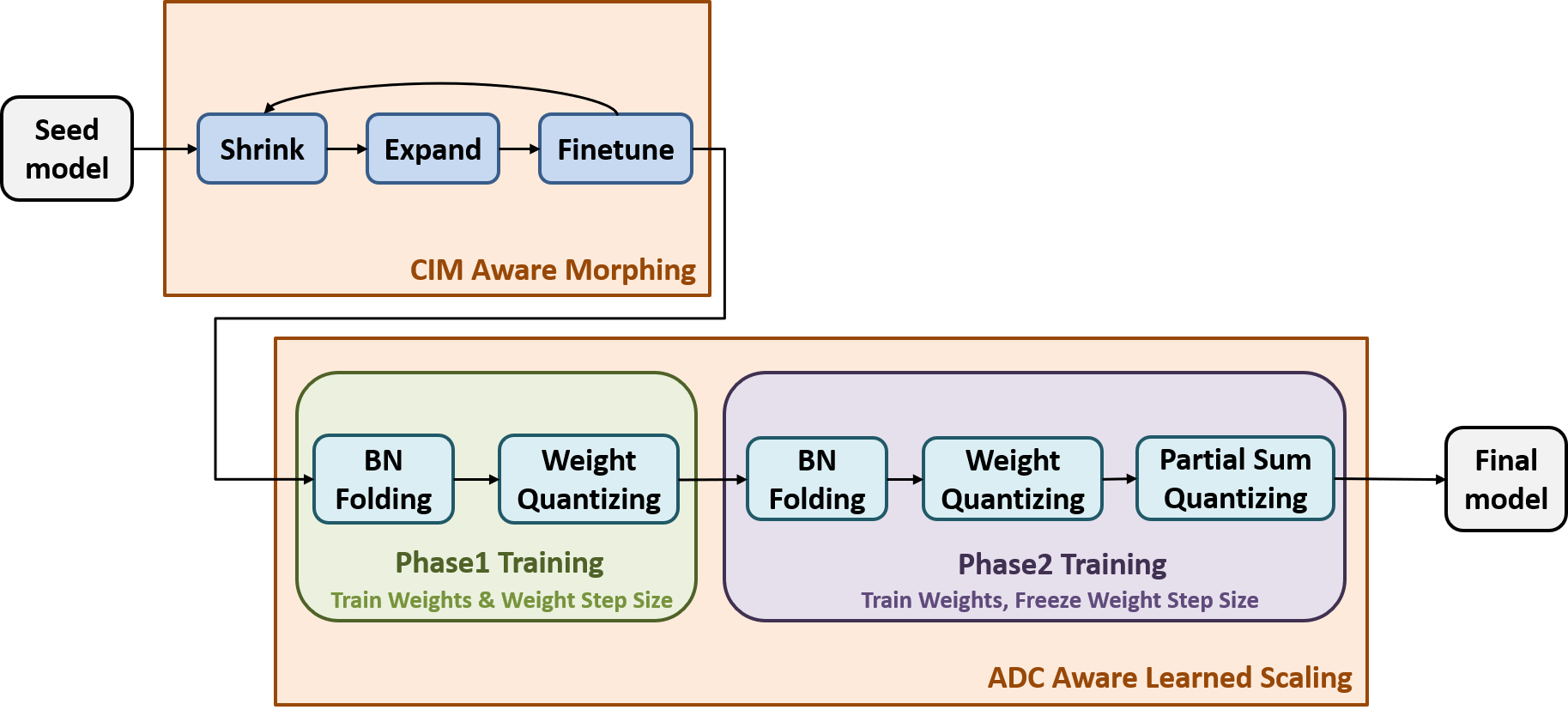}
\caption{Model adaption flow for CIM}
\label{adaption_flow}
\end{figure}

Fig.~\ref{adaption_flow} outlines the overall model adaptation flow, consisting of two stages: CIM Aware Morphing to align models with macro size, and ADC Aware Learned Scaling to scale weights based on quantization precision of both weights and ADC. CIM Aware Morphing adapts MorphNet~\cite{gordon2018morphnet} for CIM by adjusting channel numbers to fit macro size constraints like numbers of bitlines and wordlines instead of the model size or FLOPs in the original MorphNet. This iterative adjustment, typically converging in about three iterations, ensures that the model meets accuracy and resource requirements.

After roughly determining the model's shape and size, the next step involves quantizing the weights and partial sums according to the CIM weight cell's bit width, ADC precision, and ADC step size. ADC Aware Learned Scaling focuses on quantization-aware training in two steps: 
\begin{itemize}
    \item Quantization-aware training for the weights, including training the quantization step size to minimize quantization errors of weights, and
    \item Quantization-aware training for partial sums.
\end{itemize}

With these processing, the final model not only benefits from reduced redundancy through model morphing, which eliminates unnecessary filters and computations, but also addresses quantization errors through quantization-aware training, mitigating any significant accuracy drops caused by weight and partial sum quantization.

\subsection{Stage 1: CIM Aware Morphing}
CIM Aware Morphing, based on MorphNet~\cite{gordon2018morphnet},  adapts the number of channels in convolutional layers to account for the constraints of wordline and bitline quantities in CIM macros by iteratively shrinking and expanding layers within a predefined architecture. In the shrinking phase, it prunes each layer based on sparsity, varying the pruning ratio across layers. During the expansion phase, layers are proportionally scaled up according to predefined constraints, focusing on reducing computational complexity or parameter count. This targeted approach can efficiently optimize network structures without extensive architectural redesign or architectural search.

\begin{figure}[H]
\centering
\includegraphics[width=0.25\linewidth]{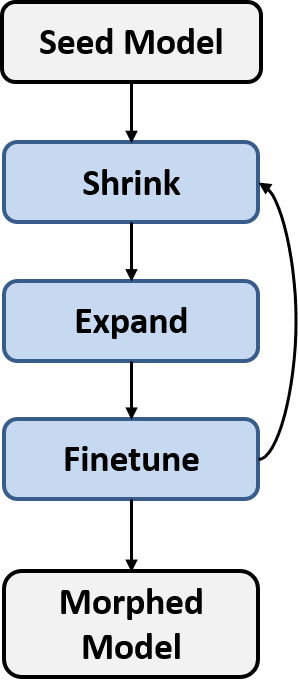}
\caption{Model morphing flow}
\label{morphing_flow}
\end{figure}

The details of the method are described below.
In the "Shrinking Stage" of a deep learning network, the loss function for channel pruning consists of two parts: the cross-entropy loss $L_{CE}(\theta)$ and the regularization term $\lambda F(\theta)$, as shown in Eq. \ref{loss_function}, where $\lambda$ is a hyper-parameter that controls the weight of the regularization term, and $\theta$ represents the model parameters. 
\begin{equation}
Loss(\theta)=L_{CE}(\theta)+\lambda F(\theta)\label{loss_function}
\end{equation}

To minimize redundancy, a regularization term related to parameter count is designed as in MorphNet \cite{gordon2018morphnet} to identify redundant parameters (see Eq. \ref{param_penalty_reformed}). The convolution filter dimensions are denoted as $x$ and $y$. Filter importance is determined by the $\gamma$ of the BN layer, with small $\gamma$ values being zeroed out to prune unimportant filters. After pruning, the remaining input and output channels, denoted as $A_{L}$ and $B_{L}$, correspond to the number of non-zero weights in the preceding and subsequent BN layers. The pruned parameter count is then calculated by multiplying $A_{L}$ and $B_{L}$ with $x$ and $y$. Here, $I_{L}$ and $O_{L}$ represent the number of input and output channels for convolution layer $L$, while $\gamma_{L-1}$ and $\gamma_{L}$ denote the BN weights before and after the convolutional layer $L$, respectively.
\begin{equation}
F(layer L)=x \times y \times (A_{L}\sum_{i=1}^{O_{L}}|\gamma_{L},i|+B_{L}\sum_{j=1}^{I_{L}}|\gamma_{L-1},j|) \label{param_penalty_reformed}
\end{equation}

To address CIM macro size constraints and identify redundancy, we use parameter count as a regularization term when adjusting channels. This approach targets deeper layers, which typically contain more redundant parameters, helping to maintain model accuracy during compression. 

For the "Expanding Phase",  it is not possible to derive the expansion ratio for CIM macros directly using an equation as with parameter expansion ratios due to the array-based structure of CIM macros. Therefore, we first list the constraint equations for the model's expansion ratio in the CIM macro as follows:
\begin{align}
    &\lceil \frac{3\times kernel\_size^2}{wordlines}\rceil \times round(C_{1}\times R) \\
    & +\sum_{i=1}^{n-1}[\lceil \frac{round(C_{i}\times R)}{channels_{per\_bl}} \rceil \times round(C_{i+1}\times R)]\leq target_{bl}
\end{align}
\begin{equation}
    channels_{per\_bl}=\lfloor \frac{wordlines}{kernel\_size^2} \rfloor
\end{equation}
Where $R$ is the desired expansion ratio, $n$ is the total number of convolutional layers, $C_{i}$ is the number of output channels in the i-th convolutional layer, and $channels_{per\_bl} $ represents the maximum number of input channels that a single bitline can accommodate.

Since solving the above inequality is very complex, we use exhaustive search here. By incrementing the ratio from 1 by 0.001 until the condition is no longer satisfied, we can find the desired expansion ratio. Additionally, only one exhaustive search is needed per morphing process, making the search very efficient. Note that the expansion ratio is applied proportionally across all layers, not a separate ratio for each layer. This makes the optimization a simple one-dimensional search for a single scalar value.

\subsection{Stage 2: ADC Aware Learned Scaling}

\begin{figure}[H]
\centering
\includegraphics[width=1\linewidth]{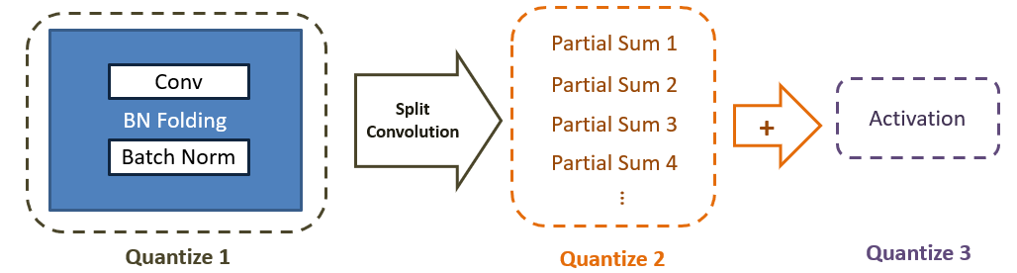}
\caption{Quantization type for models mapped to the CIM macro}
\label{quantize_type}
\end{figure}

Based on the above model adjustment, the next steps involve two rounds of quantization-aware training as shown in Fig.~\ref{quantize_type}. First, we combine convolutional and BN weights and quantize them to 4 bits to fit within a 4-bit weight macro. Second, partial sum quantization is applied to obtain the final quantized model. 

A convolution layer undergoes three types of quantization:
\begin{enumerate}
    \item \textbf{Weight Quantization}: Here, BN weights and convolutional weights from the morphed model are combined and quantized to 4 bits according to the precision of the weight cells in the CIM macro.
    \item \textbf{Partial Sum Quantization}: The partial sums are quantized to 5 bits based on the precision of the given ADC.
    \item \textbf{Activation Quantization}: This is included in the original seed model and will be quantized to 4 bits based on the DAC precision.
\end{enumerate}

\subsubsection{Phase-1: Weight Quantization Training}
\begin{figure}[H]
\centering
\includegraphics[width=0.65\linewidth]{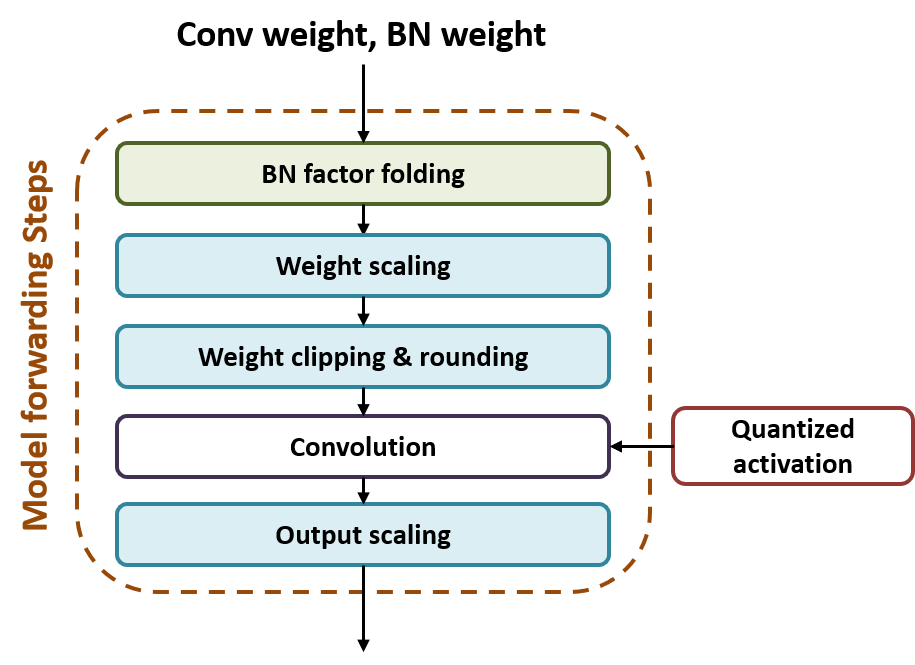}
\caption{Forwarding flow Phase1 training}
\label{phase1}
\end{figure}

Fig.~\ref{phase1} illustrates the Phase-1 weight quantization process for the model. During the forward computation of the model training, we reduce the number of parameters by combining the BN parameters with the convolutional kernel weights. These combined weights are then scaled by dividing them by the corresponding weight quantization step size, followed by clipping and rounding based on the weight bit-width. After performing the convolution with quantized activations, the results are scaled back by multiplying with the scaling factors. 

In the above process, the step size of weight quantizationis learned by the LSQ method\cite{esser2019learned}. The weight quantization equation is presented in Eq.~\ref{weight_quantization}. Here, $W$ represents the weight, $S_{W}$ is the weight quantization step size, and $-Q_{N}$ and $Q_{P}$ represent the minimum and maximum clipping values, respectively. These values are related to the number of bits being quantized. For instance, if quantizing to $n$ bits, then $Q_{N}=Q_{P}=2^{n-1}-1$. This process allows the quantization error to be reflected in floating-point representation.
\begin{equation}
    output=[round(clip(\frac{W}{S_{W}}, -Q_{N}, Q_{P}))]\ast Input\times S_{W}
    \label{weight_quantization}
\end{equation}

For our target macro, to produce 4-bit weights, the weights are first divided by $S_{W}$ for scaling (where $S_{W}$ is the weight quantization step size, typically less than 1). Then, based on the maximum and minimum values of the stored weight, the weights are clipped and rounded to obtain 4-bit weights that can be stored in the CIM macro. After performing convolution in the CIM macro with the 4-bit quantized weights, the output is multiplied by $S_{W}$ to scale it back down.

\begin{figure}[H]
\centering
\includegraphics[width=1\linewidth]{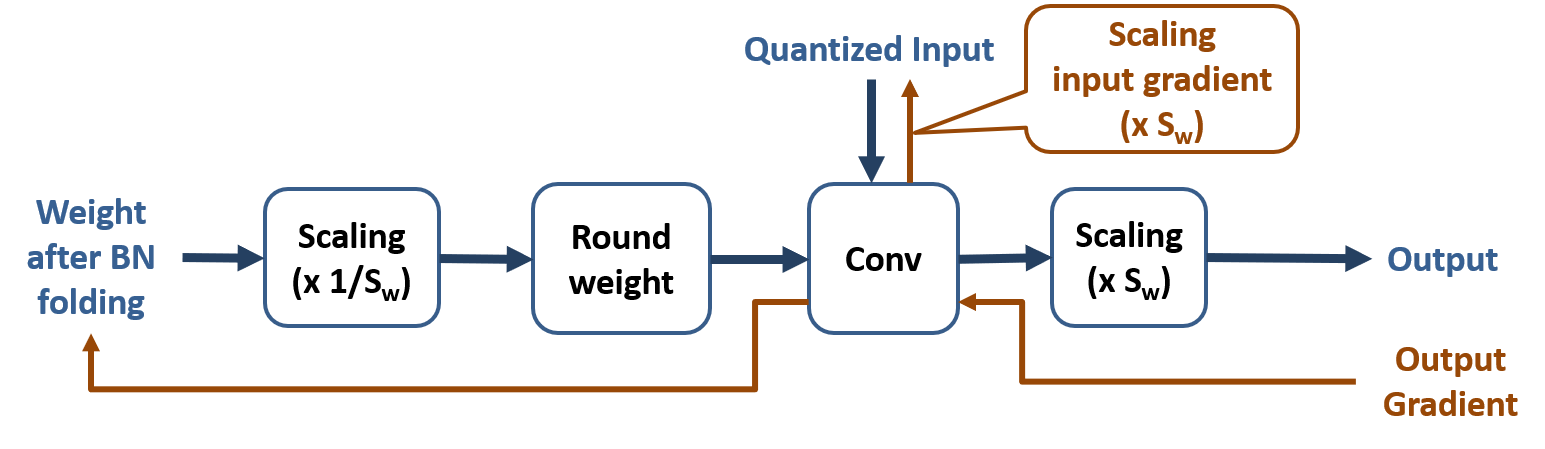}
\caption{Forward and backward data flow weight quantization}
\label{QW_backward}
\end{figure}

During the Phase-1 training, we optimize the BN and convolution weights, along with the quantization step size $S_{W}$. The goal is to complete BN weight folding and quantize the weights, as detailed in Fig.~\ref{QW_backward}.

In the backward pass, gradient computation bypasses scaling and non-differentiable rounding to maintain stability. The straight-through estimator (STE) is applied during rounding skips: gradients exceeding the clipping range are set to zero, while those within the range pass through unchanged. Additionally, since amplified weights and output gradients are used to compute input gradients, these input gradients are inversely scaled down according to the weight amplification.

\subsubsection{Phase-2: Partial Sum Quantization Training}

\begin{figure}[H]
\centering
\includegraphics[width=1\linewidth]{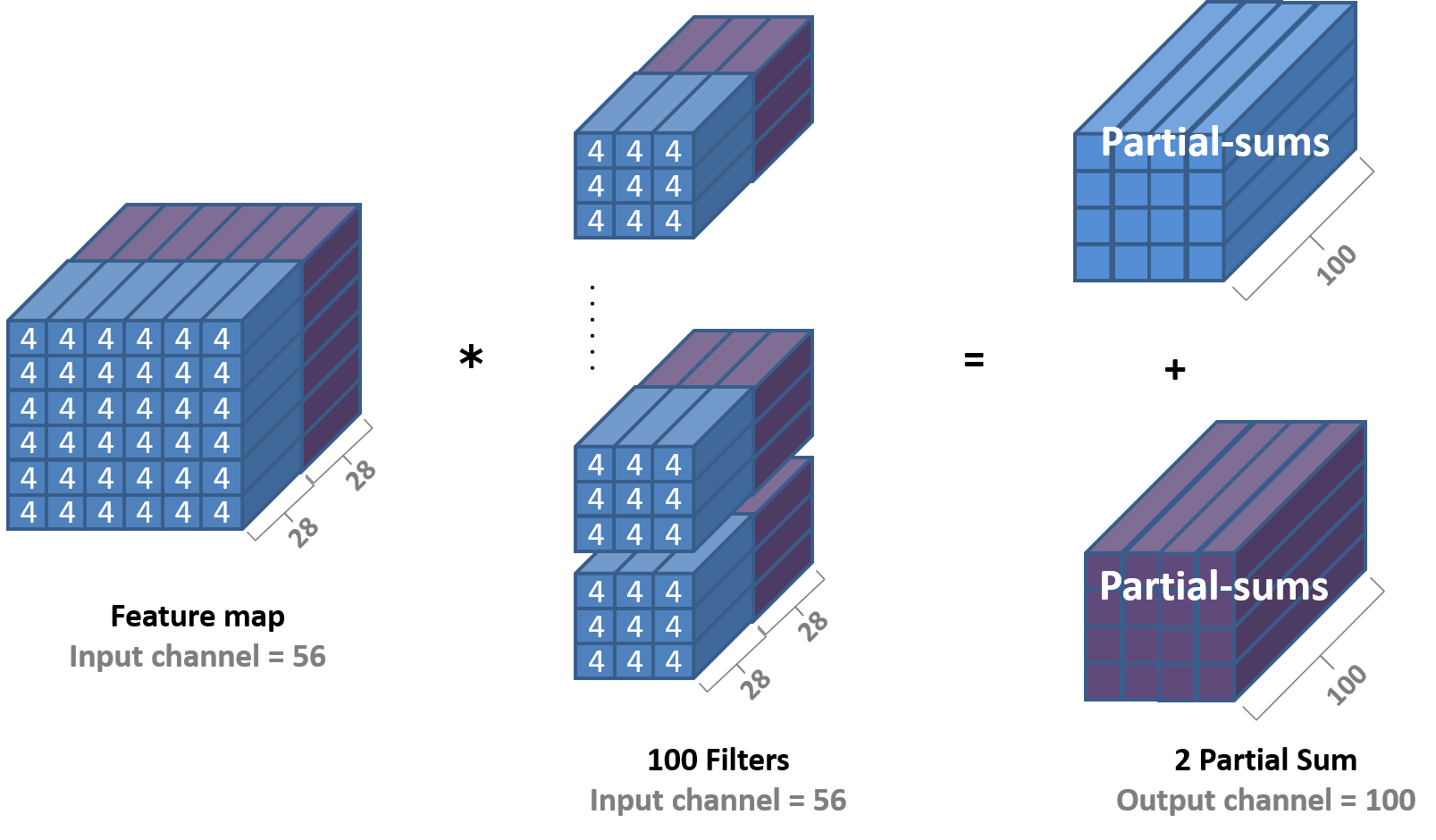}
\caption{Partial Sum Formation}
\label{psum_forming}
\end{figure}

Due to limited wordlines, larger convolutions must be processed in segments, leading to accumulated ADC quantization errors with each partial sum. To mitigate this, we incorporate partial sum quantization during Phase-2 training to simulate ADC behavior, which helps the model adapt to the quantization process. For example, as shown in Fig.~\ref{psum_forming}, with 256 wordlines, a 3x3 kernel can accommodate up to 28 input channels per bitline, requiring additional channels to be assigned to another bitline. Therefore, for a feature map and filter with 56 input channels, we divide them into two groups—denoted with blue and purple in the figure. The blue feature map convolves with the blue filters, while the purple feature map convolves with the purple filters, resulting in two partial sums that can be added point by point to obtain the final result.

\begin{figure}[H]
\centering
\includegraphics[width=0.65\linewidth]{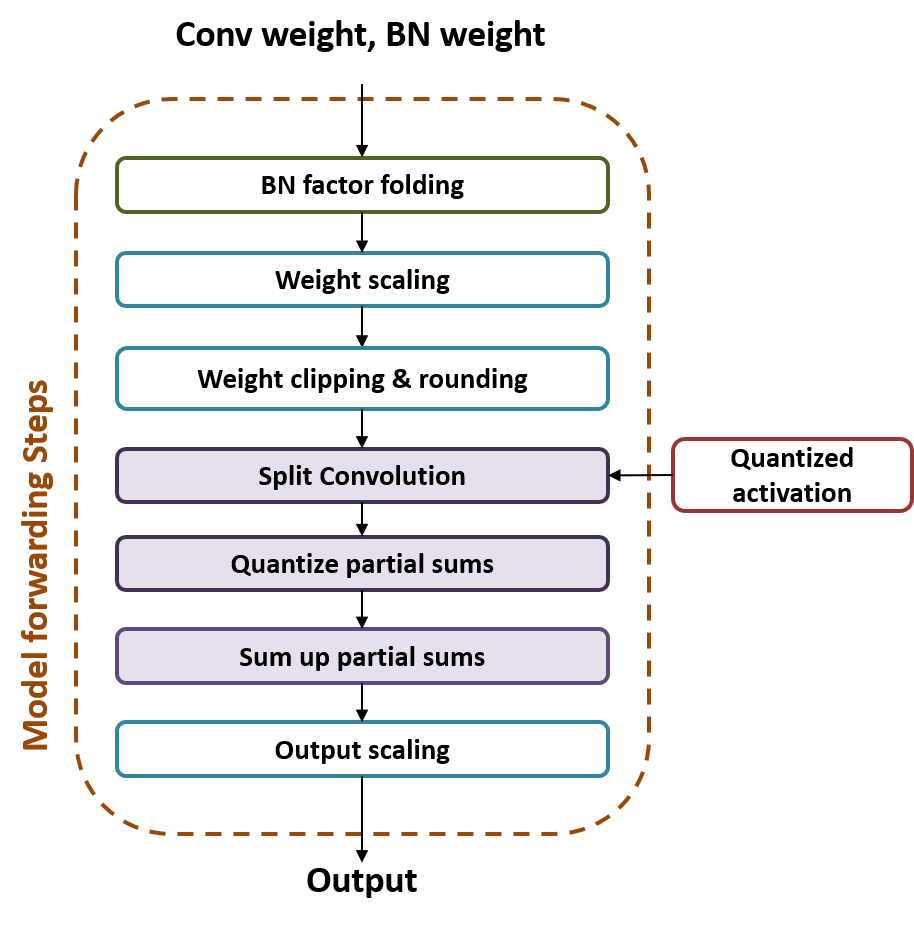}
\caption{Forwarding flow of Phase2 training}
\label{phase2}
\end{figure}

Fig.~\ref{phase2} illustrates the forwarding flow of the Phase-2 training. Compared to the Phase-1, the Phase-2 includes additional steps for the segmented convolution, quantization of partial sums, and summation of partial sums. The model output from the Phase-1 training serves as the baseline model for the Phase-2 training.

Since the Phase 2 training involves the quantization of partial sums, even minor variations in $S_{W}$ can directly affect the size of the 4-bit quantized weights if $S_{W}$ is not fixed. This, in turn, can cause significant fluctuations in the partial sums, hindering model convergence. Therefore, in the Phase 2 training, $S_{W}$ is fixed, and the BN and convolution weights are trained to adapt to the partial sum quantization.

By slightly modifying Eq. \ref{weight_quantization}, we obtain the partial sum quantization formula, as shown in Eq. \ref{psum_quantization}. This formula primarily incorporates the ADC step size and sets the  maximum and minimum clipping values according to the ADC precision, represented as $-Q_{N_{ADC}}$ and $Q_{P_{ADC}}$.

\begin{multline}
output = round\left( clip\left( \frac{Qw \cdot Input}{S_{ADC}}, \right.\right. \\
\left.\left. -Q_{N_{ADC}}, Q_{P_{ADC}} \right) \right) \cdot S_{W} \cdot S_{ADC}
\label{psum_quantization}
\end{multline}

\begin{equation}
    Qw=[round(clip(\frac{W}{S_{W}}, -Q_{N}, Q_{P}))]
    \label{weight_quantization_2}
\end{equation}

\begin{figure}[H]
\centering
\includegraphics[width=1\linewidth]{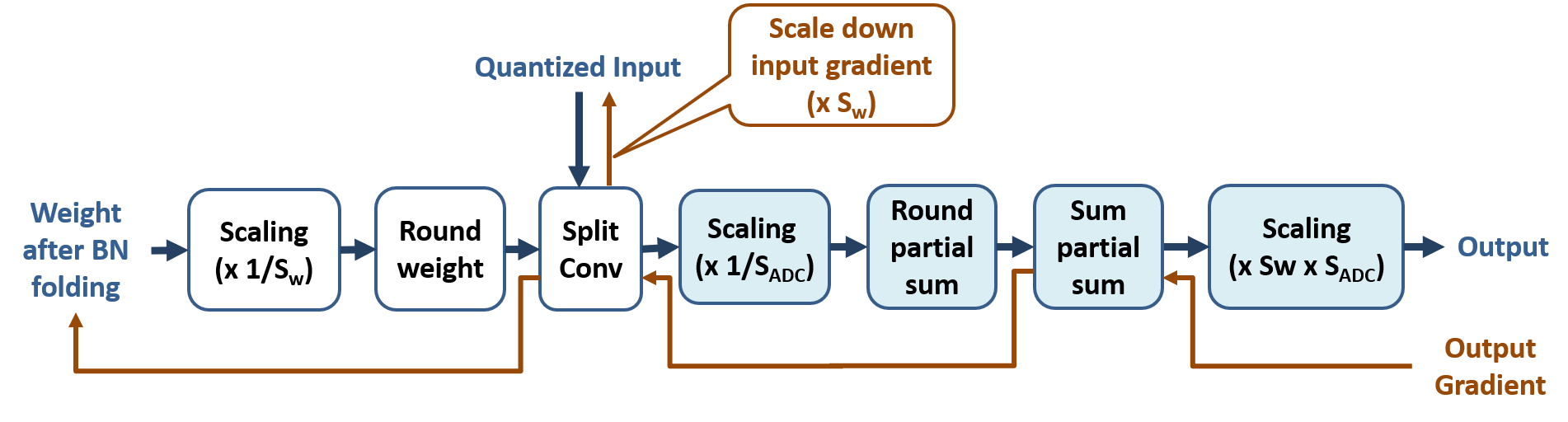}
\caption{Forward and backward data flow partial sum quantization}
\label{QP_backward}
\end{figure}

During the Phase-2 training process, only the BN and convolution weights are trained. The main goal is to adapt the weights to the quantization of partial sums. The detailed forward and backward methods are shown in Fig.~\ref{QP_backward}.

Compared to the Phase 1 training, the Phase 2 includes scaling the partial sums according to the ADC step size, followed by rounding and summing. Finally, the scaling effect of the ADC step size is inversely scaled back at the output.

In the backward pass, the gradient computation similarly skips all scaling and non-differentiable rounding operations to ensure that the gradients do not experience sudden scaling up or down, thus maintaining stability.

Finally, the trained 4-bit weights can be directly used in the CIM macro for convolution operations with 4-bit inputs. After each convolution, the output only needs to be scaled by the product of the weight step size $S_{W}$ and ADC step size $S_{ADC}$. For further simplification, this product can be approximated as a power of two, allowing the output to be adjusted with a simple digital shift operation. 

\section{Experimental Results}
\subsection{Experimental Setup}
The experimental settings for our model training are shown below. We adopt the ADAM optimizer for all trainings. The seed models used in model morphing are trained with the learning rate at 0.01 over 2000 epochs. The CIM aware morphing phase uses the learning rate at 0.05 over 100 epochs for the shrinking stage, and the learning rate at 0.01 over 100 epochs for the following fine-tuning stage, respectively. The ADC aware learning scaling adopts the learning rate at 0.001 with 100 epochs at the phase-1, and the learning rate at 0.01 over 300 epochs at the phase-2, respectively.

\subsection{Analysis of Parameter Selection for the Model Morphing}
The CIM aware model morphing has shown how to morph the model under the macro constraints. However, how to select the ratio of compression and expansion is crucial for model performance and hardware utilization of the CIM macro.

As an example to show the effect of compression ratio, Table \ref{compression_limit} shows the accuracy of models with different compression ratios after being expanded to the same parameter count and fine-tuned. The baseline model has 9.218M parameters and an accuracy of 90.71\%. The target for expansion is set at 50\% of the baseline parameters, totaling 4.609M. This table shows that excessive compression (e.g. \( \text{pruning\ ratio} > 0.9 \)) will decrease performance due to a loss of important features. However, insufficient compression (e.g. \( \text{pruning\ ratio} < 0.1 \)) limits the effectiveness of expansion and thus decreases performance as well.
In addition to performance concerns, these ratios also lead to different macro usage due to macro constraints.

\begin{table}[H]
    \centering
    \caption{Model compression limit}
    \begin{tabular}{|c||c|c|}
    \hline
    \textbf{Parameters} & \textbf{Parameters} & \textbf{Accuracy} \\
    \textbf{(Pruned)} & \textbf{(Expanded)} &  \\
    \hline
    0.429M & 4.611M & 87.66\% \\
    \hline
    0.501M & 4.607M & 88.94\% \\
    \hline
    0.691M & 4.608M & 89.70\% \\
    \hline
    1.014M & 4.605M & 90.70\% \\
    \hline
    1.262M & 4.609M & 90.90\% \\
    \hline
    1.993M & 4.609M & 90.90\% \\
    \hline
    2.445M & 4.604M & 90.70\% \\
    \hline
    2.848M & 4.610M & 90.76\% \\
    \hline
    3.791M & 4.607M & 90.62\% \\
    \hline
    4.049M & 4.610M & 90.32\% \\
    \hline
    \end{tabular}
    \label{compression_limit}
\end{table}

Table \ref{different_macro_usage} shows the accuracy differences after expansion and fine-tuning for two models with varying macro utilization rates by a grid search on the parameters of the model morphing flow. In this table, the top two rows are the best and worst macro usage when $\lambda=5E-8$. The bottowm two rows are the best and worst macro usage when $\lambda=3E-8$.
The baseline model has 9.218M parameters and an accuracy of 90.71\%. The target for model expansion is set at 8192 bitlines and 256 wordlines, using the ADAM optimizer for both compression and fine-tuning. During the 150-epoch compression phase, the learning rate is 0.05, and $\lambda$ is gradually increased from 0 over the first 100 epochs before being fixed for the last 50 epochs. Compressed models with the highest and lowest macro usage are compared. After expansion, models are fine-tuned for 300 epochs at a learning rate of 0.01.

\begin{table}[H]
    \centering
    \caption{Result of different CIM macro usage model for the VGG-9 model on CIFAR-10.}
    \begin{tabular}{|c|c|c|c|}
    \hline
    \textbf{Parameters} & \textbf{Parameters} & \textbf{Macro Usage} &  \textbf{Accuracy} \\
    \textbf{(Pruned)} & \textbf{(Expanded)} &  & \\
    \hline
    1.154M & 1.960M & 93.46\% & 91.16\% \\
    \hline
    1.203M & 1.867M & 88.53\% & 90.97\% \\
    \hline
    1.255M & 1.929M & 92.00\% & 91.01\% \\
    \hline
    1.413M & 1.833M & 87.41\% & 90.88\% \\
    \hline
    \end{tabular}
    \label{different_macro_usage}
\end{table}

Table \ref{compression_limit} shows that model performance declines when the compression ratio falls below a certain threshold, e.g. 0.1 in Table \ref{compression_limit}. Therefore, below this threshold, it's crucial to select a model that retains feature representation rather than focusing solely on CIM macro utilization after expansion. Thus, if the target macro size is less than 0.1 times the baseline model’s parameter count, it's better to choose a model with higher accuracy during compression. In contrast, if the target macro size exceeds 0.1 times the baseline count, there’s less risk of losing feature representation, making it acceptable to select a model with higher CIM macro utilization. This strategy can help achieve higher accuracy through resource reallocation.

\subsection{End-to-End Performance}
This subsection present the main results for latency, accuracy, and compression across different models.
\subsubsection{Settings}
To show the effectiveness of the proposed approach, the model adaption have been applied to different models, VGG9, VGG16, and ResNet18, as shown in Tables \ref{comprehensive_results_VGG9} to \ref{comprehensive_results_ResNet18}, tailored to the constraints of four CIM macros, focusing on wordline and bitline limitations, as well as quantization restrictions for weight cells and ADCs.
These tables display accuracy based on CIFAR-10 test performance, where BLs denotes the number of bitlines in the CIM macro architecture (256 wordlines), and MACs represents the multiply-accumulate operations required for inference (equivalent to ADC activations). The baseline model features 4-bit quantized activations and was trained on CIFAR-10 for 2000 epochs. Four models are created under varying bitline constraints, each undergoing three morphing rounds: a 150-epoch compression phase and a 300-epoch fine-tuning phase, both utilizing the ADAM optimizer (with learning rates of 0.05 and 0.01, respectively).

In the tables, Morphed Model Accuracy indicates the model's accuracy after compression. The Phase-1 Training shows accuracy after batch normalization (BN) folding and 4-bit weight quantization, while the Phase2 Training reflects accuracy after further 5-bit partial quantization. 
The partial sum storage and latency reduction presents model weights allocated in a CIM macro with 256 bitlines and wordlines, each featuring 4-bit weight cells. Due to limited wordlines, 5-bit partial sums are generated, necessitating additional storage, with Partial Sum Storage indicating the maximum space required for these sums. Loading Weight Latency estimates the clock cycles needed to load weights; a CIM macro would require 256 cycles for this process.
Lastly, Computing Latency denotes the clock cycles required for model inference. Convolution filters are divided into smaller chunks that convolve with input channels sequentially, necessitating multiple passes through the wordlines. With only 64 ADCs available (4 bitlines per ADC), exceeding 64 simultaneous computations requires additional passes. The table provides the clock cycles needed for a CIM macro to perform model inference.

\subsubsection{Results}
Tables \ref{comprehensive_results_VGG9} to \ref{comprehensive_results_ResNet18} present the results for VGG9/VGG16/ResNet18 after model morphing and weight adaptation, respectively.
VGG9 comprises 8 convolutional layers and 1 fully connected layer. VGG16 features 13 convolutional layers and 1 fully connected layer. ResNet18 has 17 convolutional layers and 1 fully connected layer. 
For simplicity, only the convolutional layers are accelerated by the CIM macros.

The model morphing results indicate that for models utilizing over 4096 bitlines (aka. more parameters), reallocating resources improves accuracy (91.33\% and 91.07\% in VGG9, 92.98\% and 92.66\% in VGG16, and 92.17\% in ResNet18) compared to the baseline, even with fewer bitlines and total MAC operations. This enhancement stems from pruning redundant filters and reallocating excess bitline resources to critical convolutional layers, resulting in more meaningful and efficient weight storage and operations within the CIM macro. 

The proposed morphing can also achieve high macro usage, up to 94.54\%, with small accuracy loss due to the CIM aware constraints. 
The macro usage for ResNet18 is lower compared to the VGG models due to the higher number of convolutional layers. Consequently, with a bitline limit of 4096, the model's accuracy is slightly declined. When the limit is reduced to 512, accuracy decreases further to just 25\% macro usage, resulting in lower accuracy than the VGG models. Additionally, as the number of parameters is decreased, quantization significantly impacts accuracy, causing an extra 3.75\% drop when the bitline limit is 512.

The proposed quantization (\textit{P1 train} and \textit{P2 train} in the tables) can achieve low accuracy loss for the bitline constraints over 4096. The quantization loss will be increased for smaller bitline constraints, which are reasonable since small model size has low tolerance to quantization effects. The tables also show that the proposed partial sum quantization (\textit{P2 train}) has introduced negligible loss compared to the weight quantization (\textit{P1 train}). 

In the tables, the partial sum storage are reduced due to model morphing except one case. With a bitline constraint of 8192, partial sum storage for VGG16 is increasesd. This occurs because the additional bitlines from pruning are allocated to earlier layers, which are critical for accuracy. These layers require more partial sum storage as their feature maps have not undergone significant pooling.  

The computing latency for all cases is reduced (26\% to 86\% for VGG9, 30\% to 89\%,3\% to 81\% for ResNet18), which is proportional to the reduction of the MACs due to the model morphing. The latency to reload weight due to the limited macro size is also reduced (79\% to 99\% for VGG9, 87\% to 99\% for VGG16 and 82\% to 99\% for ResNet18), which has higher reduction ratio than that in the computing latency due to the CIM constraint. These ratios are proportional to the reduction of the parameters and used BLs.

\begin{table*}[htb]
    \centering
    \caption{Comprehensive Results for VGG9 with Different BL Constraints}
    \scriptsize %
    \setlength{\tabcolsep}{3pt} %
    \renewcommand{\arraystretch}{1.2} %
    \begin{tabular}{|c||c|c|c|c|c|c|c|c|c|c|}
    \hline
    \textbf{BL} & \textbf{Param} & \textbf{BLs} & \textbf{MACs} & \textbf{Macro} & \textbf{Morphed Model} & \textbf{P1} & \textbf{P2} & \textbf{Partial sum} & \textbf{Load Weight} & \textbf{Computing} \\
    \textbf{Constraint} & (M) & & & \textbf{Usage} & \textbf{Acc.} & \textbf{Train} & \textbf{Train} & \textbf{Storage} & \textbf{Latency} & \textbf{Latency} \\
    \hline
    \textbf{Baseline} & 9.218 & 38592 & 724992 & - & 90.71\% & - & - & 163840 & 38656 & 14696 \\
    \hline
    \textbf{8192} & 1.971 (-79\%) & 8186 (-79\%) & 489248 (-33\%) & 93.98\% & 91.33\% (+0.62\%) & 90.01\% & 89.83\% & 133056 (-19\%) & 8192 (-79\%) & 10928 (-26\%) \\
    \hline
    \textbf{4096} & 0.924 (-90\%) & 3907 (-90\%) & 358888 (-50\%) & 88.12\% & 91.07\% (+0.36\%) & 89.77\% & 89.17\% & 107520 (-34\%) & 4096 (-89\%) & 9116 (-38\%) \\
    \hline
    \textbf{1024} & 0.210 (-98\%) & 1024 (-97\%) & 123792 (-83\%) & 80.11\% & 89.24\% (-1.47\%) & 87.58\% & 87.39\% & 41984 (-74\%) & 1024 (-97\%) & 3020 (-80\%) \\
    \hline
    \textbf{512} & 0.098 (-99\%) & 511 (-99\%) & 85756 (-88\%) & 74.77\% & 87.71\% (-3.00\%) & 85.47\% & 85.40\% & 39936 (-76\%) & 512 (-99\%) & 2108 (-86\%) \\
    \hline
    \end{tabular}
    \label{comprehensive_results_VGG9}
\end{table*}

\begin{table*}[htb]
    \centering
    \caption{Comprehensive Results for VGG16 with Different BL Constraints}
    \scriptsize %
    \setlength{\tabcolsep}{3pt} %
    \renewcommand{\arraystretch}{1.2} %
    \begin{tabular}{|c||c|c|c|c|c|c|c|c|c|c|}
    \hline
    \textbf{BL} & \textbf{Param} & \textbf{BLs} & \textbf{MACs} & \textbf{Macro} & \textbf{Morphed Model} & \textbf{P1} & \textbf{P2} & \textbf{Partial sum} & \textbf{Load Weight} & \textbf{Computing} \\
    \textbf{Constraint} & (M) & & & \textbf{Usage} & \textbf{Acc.} & \textbf{Train} & \textbf{Train} & \textbf{Storage} & \textbf{Latency} & \textbf{Latency} \\
    \hline
    \textbf{Baseline} & 14.710 & 61440 & 1443840 & - & 92.02\% & - & - & 196608 & 61440 & 31300 \\
    \hline
    \textbf{8192} & 1.983 (-87\%) & 8148 (-87\%) & 986784 (-32\%) & 94.54\% & 92.98\% (+0.96\%) & 92.73\% & 92.25\% & 245760 (+25\%) & 8192 (-87\%) & 21996 (-30\%) \\
    \hline
    \textbf{4096} & 0.952 (-94\%) & 3963 (-94\%) & 622032 (-57\%) & 90.83\% & 92.66\% (+0.64\%) & 92.49\% & 91.88\% & 174080 (-11\%) & 4096 (-93\%) & 16192 (-48\%) \\
    \hline
    \textbf{1024} & 0.203 (-99\%) & 1021 (-98\%) & 259420 (-82\%) & 77.58\% & 89.96\% (-2.06\%) & 88.66\% & 88.55\% & 106496 (-46\%) & 1024 (-98\%) & 6028 (-81\%) \\
    \hline
    \textbf{512} & 0.088 (-99\%) & 510 (-99\%) & 117408 (-92\%) & 67.07\% & 86.45\% (-5.57\%) & 83.03\% & 84.50\% & 35840 (-82\%) & 512 (-99\%) & 3532 (-89\%) \\
    \hline
    \end{tabular}
    \label{comprehensive_results_VGG16}
\end{table*}

\begin{table*}[htb]
    \centering
    \caption{Comprehensive Results for ResNet18 with Different BL Constraints}
    \scriptsize %
    \setlength{\tabcolsep}{3pt} %
    \renewcommand{\arraystretch}{1.2} %
    \begin{tabular}{|c||c|c|c|c|c|c|c|c|c|c|c|}
    \hline
    \textbf{BL} & \textbf{Param} & \textbf{BLs} & \textbf{MACs} & \textbf{Macro} & \textbf{Morphed Model} & \textbf{P1} & \textbf{P2} & \textbf{Partial sum} & \textbf{Load Weight} & \textbf{Computing} \\
    \textbf{Constraint} & (M) & & & \textbf{Usage} & \textbf{Acc.} & \textbf{Train} & \textbf{Train} & \textbf{Storage} & \textbf{Latency} & \textbf{Latency} \\
    \hline
    \textbf{Baseline} & 10.987 & 46400 & 690176 & - & 91.44\% & - & - & 65536 & 46592 & 16860 \\
    \hline
    \textbf{8192} & 1.804 (-84\%) & 8188 (-82\%) & 674344 (-2\%) & 86.01\% & 92.17\% (+0.73\%) & 91.34\% & 90.99\% & 97280 (+48\%) & 8192 (-82\%) & 16296 (-3\%) \\
    \hline
    \textbf{4096} & 0.829 (-92\%) & 4088 (-91\%) & 411848 (-40\%) & 78.77\% & 91.37\% (-0.07\%) & 90.40\% & 90.21\% & 66560 (+2\%) & 4096 (-91\%) & 12092 (-28\%) \\
    \hline
    \textbf{1024} & 0.132 (-99\%) & 997 (-98\%) & 145888 (-79\%) & 50.71\% & 86.16\% (-5.28\%) & 84.37\% & 84.68\% & 57344 (-13\%) & 1024 (-98\%) & 3940 (-77\%) \\
    \hline
    \textbf{512} & 0.033 (-99.6\%) & 512 (-99\%) & 79760 (-88\%) & 25.37\% & 81.01\% (-10.43\%) & 78.74\% & 77.26\% & 40960 (-38\%) & 512 (-99\%) & 3128 (-81\%) \\
    \hline
    \end{tabular}
    \label{comprehensive_results_ResNet18}
\end{table*}

\begin{figure}[htb]
\centering
\includegraphics[width=1.0\linewidth]{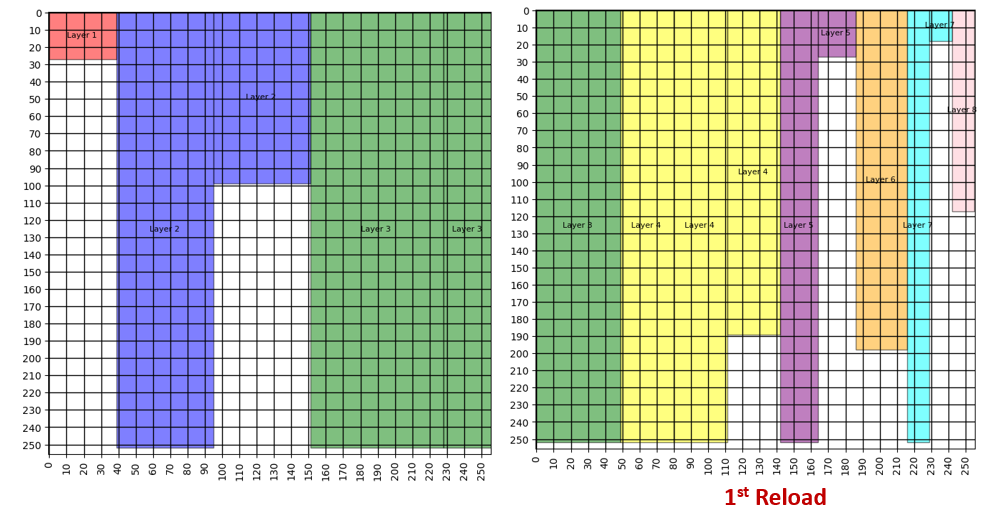}
\caption{Mapping convolution weights into a CIM macro (model: VGG9,  BL constraint: 512)}
\label{VGG9_512}
\end{figure}

\begin{figure}[htb]
\centering
\includegraphics[width=1.0\linewidth]{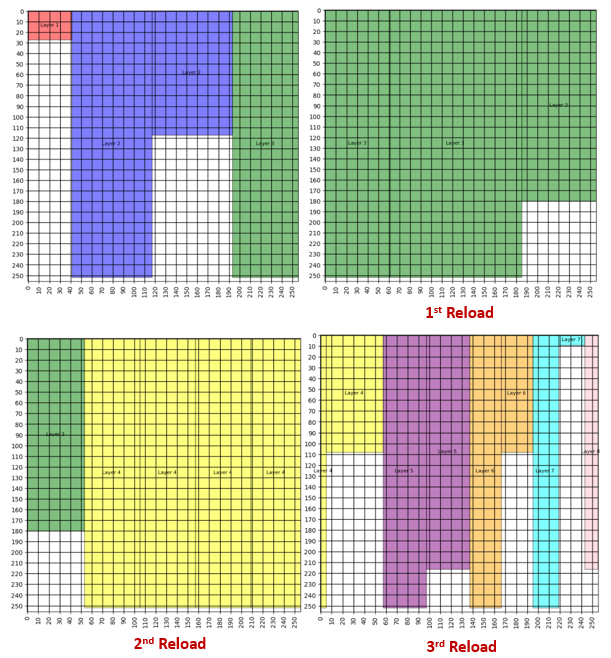}
\caption{Mapping convolution weights into a CIM macro (model: VGG9,  BL constraint: 1024)}
\label{VGG9_1024}
\end{figure}

Fig.~\ref{VGG9_512} and ~\ref{VGG9_1024} illustrate the mapping of the VGG9 model, morphed under bitline constraints of 512 and 1024, onto a 256x256 CIM macro. Different colors in the figures represent different convolutional layers. 

\begin{table*}[htbp]
\centering
\caption{Comparison table}
\begin{tabular}{|c|c|c|c|c|c|c|}
\hline
                            & E-UPQ \cite{chang2023upq}           & E-UPQ  \cite{chang2023upq}           & XPert \cite{moitra2023xpert}           & \multicolumn{3}{c|}{\textbf{This work}}                                                                                      \\ \hline
Model                       & ResNet18        & ResNet20        & VGG16            & \textbf{VGG9}             & \textbf{VGG16}            & \textbf{ResNet18}          \\ \hline

Dataset                     & CIFAR-100       & CIFAR-10        & CIFAR-10         & \textbf{CIFAR-10}         & \textbf{CIFAR-10}         & \textbf{CIFAR-10}          \\ \hline

Baseline           & 74.4\%          & 91.3\%          & 94.0\%           & \textbf{90.7\%}           & \textbf{92.0\%}           & \textbf{91.4\%} \\
accuracy           &           &           &            &            &            &             \\ \hline

Compressed         & 73.2\% & 90.5\%  & 92.46\%  & \textbf{89.17\%} & \textbf{91.88\%} & \textbf{90.21\%} \\ 
accuracy         & (-1.2\%) & (-0.8\%) & (-1.5\%) & \textbf{(-1.5\%)} & \textbf{(-0.8\%)} & \textbf{(-1.23\%)} \\ \hline

Bit (Weight/ & 1.0/8.0/4.0     & 1.1/8.0/4.0     & 8.0/4.0/5.4      & \textbf{4.0/4.0/5.0}      & \textbf{4.0/4.0/5.0}      & \textbf{4.0/4.0/5.0} \\
Activation/ADC) &      &      &       &       &       &        \\ \hline

Memory cell                 & 1 bit           & 1 bit           & 1 bit            & \textbf{4 bits}           & \textbf{4 bits}           & \textbf{4 bits}            \\ \hline

Compression ratio           & -87.50\%        & -86.30\%        & -68.41\%         & \textbf{-89.98\%}         & \textbf{-93.53\%}         & \textbf{-92.45\%}          \\ \hline

Macro usage                 & 12.50\%         & 13.70\%         & -                & \textbf{88.12\%}          & \textbf{90.83\%}          & \textbf{78.77\%}           \\ \hline

Activated wordlines         & 16              & 16              & 64               & \textbf{256}              & \textbf{256}              & \textbf{256}               \\ \hline

Pruning                     & $\checkmark$    & $\checkmark$    & $\times$         & $\checkmark$              & $\checkmark$              & $\checkmark$               \\ \hline

Adjustable    & $\times$        & $\times$        & $\times$         & $\checkmark$              & $\checkmark$              & $\checkmark$    \\
after pruning    &         &         &          &               &               &                \\ \hline

ADC aware          & $\times$        & $\times$        & $\times$         & $\checkmark$              & $\checkmark$              & $\checkmark$   \\
training          &         &         &         &               &               &               \\ \hline

\end{tabular}
\label{comparison_table}
\end{table*}

\subsection{Comparisons with Other Approaches}
Table \ref{comparison_table} compares three model adaptation methods using a model with a 4096-bitline constraint. E-UPQ \cite{chang2023upq} employs mixed precision (8, 4, 2, 1, 0) for weights, resulting in an average precision around 1 due to extensive pruning. It uses a 16x16 operation unit (OU), activating 16 wordlines at a time, and achieves about 87\% weight reduction.  XPert \cite{moitra2023xpert} uses full floating-point operations in its baseline model, while its compressed model adopts mixed precision for activations and ADCs, averaging 4.0 and 5.4 bits, respectively, with weights fixed at 8 bits. It activates 64 wordlines simultaneously, reducing parameters by 68.41\% with 92.46\% accuracy.

Compared to the previous approaches, our method begins with 4-bit quantized activations and floating-point weights, achieving over 90\% compression through morphing and quantizing while maintaining comparable accuracy. This approach outperforms other methods in three aspects:

\begin{enumerate}
    \item \textbf{Parallelism}: By using 4-bit parallel input and activating 256 wordlines simultaneously, our method leverages ADC-aware training to handle higher quantization errors from concurrent operations. This achieves up to 64x speedup compared to E-UPQ and 16x compared to XPert.
    \item \textbf{CIM Macro Utilization}: Our method achieves nearly 90\% utilization in VGG9 and VGG16, and 78.77\% in ResNet18 with a 4096-bitline constraint, compared to just 13\% in E-UPQ. This is due to directly pruning inefficient weights instead of storing them, making more efficient use of CIM macro space. 
    \item \textbf{Compression Rate}: Through pruning and resource reallocation, our method improves accuracy and compensates for quantization-induced losses, achieving over 90\% model compression. 
\end{enumerate}

\label{session:Conclusion}

\section{Conclusion}

CIM brings the benefits of highly parallel computation and low power consumption but suffers from throughput and performance bottlenecks due to extra weight loading for limited memory array size and ADC quantization errors for partial sum. Addressing this problem, this paper has presented a two-stage process to adapt models to CIM constraints. The first stage compresses and reallocates the weights to maximize macro utilization and minimize weight loading while retaining accuracy under the CIM array size constraints. The second stage quantizes the model with the learning quantization step size and ADC aware training to reduce the impact of quantization errors for partial sum accumulation. Compared to the previous approaches, the presented method achieves higher macro utilization, up to 90\%, higher compression ratio, up to 93\%, and more activated wordlines, up to 256, with lower accuracy loss.

\bibliographystyle{IEEEtran}

% \bibliography{bib/ieeeBSTcontrol,bib/thesis}
% Generated by IEEEtran.bst, version: 1.14 (2015/08/26)

\begin{IEEEbiography}[{\includegraphics[width=1in,height=1.25in,clip,keepaspectratio]{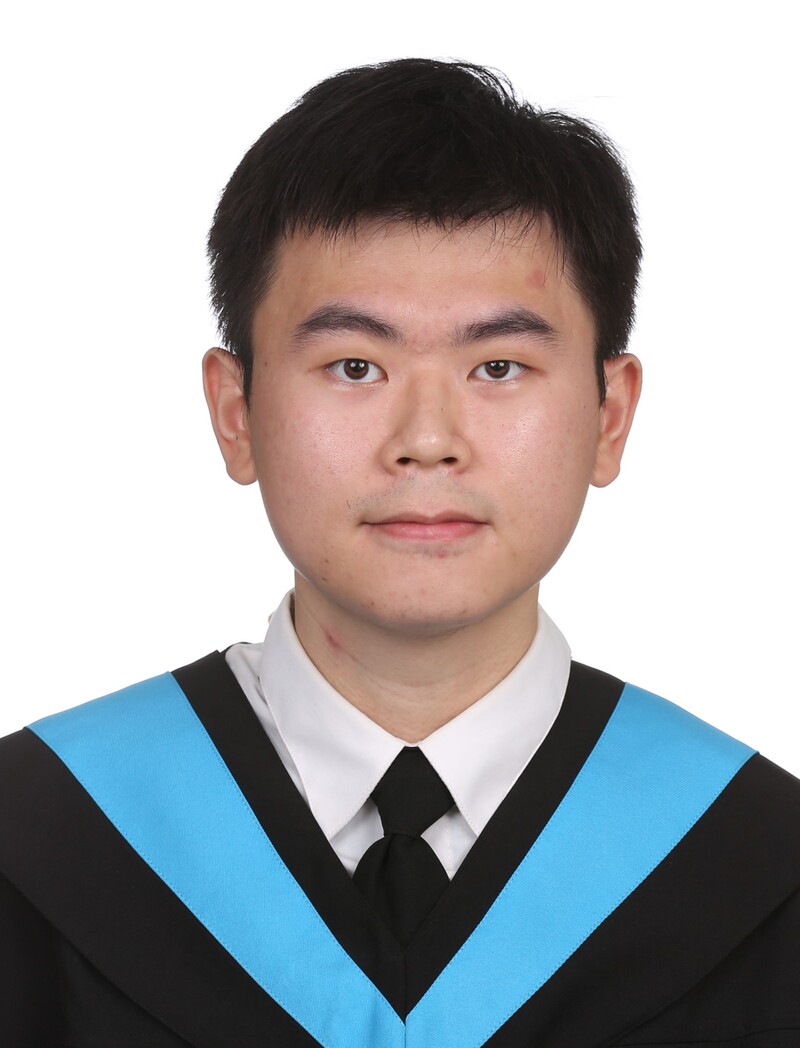}}]{Ming-Han Lin}
received the M.S. degree in electronics engineering from the National Yang Ming Chiao Tung University, Hsinchu, Taiwan, in 2024. He is currently working in the NovaTek, Hsinchu, Taiwan. His research interest includes deep learning and Computing-In-Memory.

\end{IEEEbiography}

\begin{IEEEbiography}[{\includegraphics[width=1in,height=1.25in,clip,keepaspectratio]{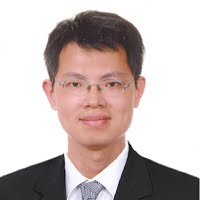}}]{Tian-Sheuan Chang}
	(S’93–M’06–SM’07)
	received the B.S., M.S., and Ph.D. degrees in electronic engineering from National Chiao-Tung University (NCTU), Hsinchu, Taiwan, in 1993, 1995, and 1999, respectively. 
	
	From 2000 to 2004, he was a Deputy Manager with Global Unichip Corporation, Hsinchu, Taiwan. In 2004, he joined the Department of Electronics Engineering, NCTU (as National Yang Ming Chiao Tung University (NYCU) in 2021), where he is currently a Professor. In 2009, he was a visiting scholar in IMEC, Belgium. His current research interests include system-on-a-chip design, VLSI signal processing, and computer architecture.
	
	Dr. Chang has received the Excellent Young Electrical Engineer from Chinese Institute of Electrical Engineering in 2007, and the Outstanding Young Scholar from Taiwan IC Design Society in 2010. He has been actively involved in many international conferences as an organizing committee or technical program committee member.
\end{IEEEbiography}
\end{document}